# A Modified VGG19-Based Framework for Accurate and Interpretable Real-Time Bone Fracture Detection

1st Md. Ehsanul Haque
*Department of CSE*
*East West University*
Dhaka, Bangladesh
ehsanulhaquesohan758@gmail.com

2nd Abrar Fahim
*Dept. of EEE*
*Islamic University of Technology*
Dhaka, Bangladesh
abrarfahim8@iut-dhaka.edu

3rd Shamik Dey
*Dept of EEE*
*Ahsanullah University of Science and Technology*
Dhaka, Bangladesh
shamikdey7@gmail.com

4th Syoda Anamika Jahan
*Dept of MPE*
*Islamic University of Technology*
Dhaka, Bangladesh
anamikajahan@iut-dhaka.edu

5th S. M. Jahidul Islam
*Dept. of CSM*
*Bangladesh Agricultural University*
Mymensingh, Bangladesh
jahidul.ict@bau.edu.bd

6th Sakib Rokoni
*Dept of CSE*
*Brac University*
Dhaka, Bangladesh
sakibrokoni61@gmail.com

7th Md Sakib Morshed
*Department of CSE*
*East West University*
Dhaka 1212, Bangladesh
morshedsakib41@gmail.com

*Abstract*—Early and accurate detection of the bone fracture is paramount to initiating treatment as early as possible and avoiding any delay in patient treatment and outcomes. Interpretation of X-ray image is a time consuming and error prone task, especially when resources for such interpretation are limited by lack of radiology expertise. Additionally, deep learning approaches used currently, typically suffer from misclassifications and lack interpretable explanations to clinical use.In order to overcome these challenges, we propose an automated framework of bone fracture detection using a VGG-19 model modified to our needs. It incorporates sophisticated preprocessing techniques that include Contrast Limited Adaptive Histogram Equalization (CLAHE), Otsu's thresholding, and Canny edge detection, among others, to enhance image clarity as well as to facilitate the feature extraction.Therefore, we use Grad-CAM, an Explainable AI method that can generate visual heatmaps of the model's decision making process, as a type of model interpretability, for clinicians to understand the model's decision making process. It encourages trust and helps in further clinical validation. It is deployed in a real time web application, where healthcare professionals can upload X-ray images and get the diagnostic feedback within 0.5 seconds. The performance of our modified VGG-19 model attains 99.78% classification accuracy and AUC score of 1.00, making it exceptionally good. The framework provides a reliable, fast, and interpretable solution for bone fracture detection that reasons more efficiently for diagnoses and better patient care.

*Index Terms*—Bone Fracture Detection, Healthcare AI , Medical Image Analysis, Diagnosis Using Machine Learning, Clinical Decision Support.

## I. Introduction

Bone fractures are a common medical condition that occur as the result of accidents, falling or degenerative disease such as osteoporosis [1] [2]. Diagnosis of fractures are usually done by radiographic imaging, that include x-rays which are well suited for instant and easy accessibility. The manual interpretation of these images by radiologists is time consuming as is error prone particularly for the less obvious fractures or poor picture quality [3]. Deep learning has arisen as one of the core agents of automation of medical image analysis in the recent years with advance gains in diagnostic precision and erosion of manual work [4]–[6]. In addition to the above, the use of machine learning (ML) and explainable artificial intelligence (XAI) has improved these systems more than ever as now they allow not only high-performance analysis but also transparency and interpretability in model decisions [7], [8].

In spite of progress, the current fracture detection models are limited. However, most of the existing approaches sacrifice the interpretability of the model while focusing on the accuracy without considering the need for it in clinical environments. At the same time, models are affected by variations in image quality, noise and lighting conditions, little systems have been designed for real time deployment or integration into clinical workflows. This also points out a gap in current research, as there is currently no robust solution, which lacks interpretability and deployment in real time, for detecting real time bone fracture.

To fill this gap, we bring forward a modified VGG-19 based deep learning frame work which is accurate, explainable and fast. Our methodology involves advanced preprocessing techniques: Contrast Limited Adaptive Histogram Equalization (CLAHE), Otsu's thresholding and Canny edge detection so as to achieve better quality and emphasis on important features in the images. In order to make our predictions as transparent as possible, we combine Grad-CAM, an explainable AI technique that creates image regions responsible for the model's decisions. Using this framework, we implement it as a real time web app where clinicians can upload images x-rays which they receive back with diagnostic feedback in 0.5 seconds or less.

The remainder of this paper is organized as follows: : Section II reviews related work while Section III describes methodology. Section IV presents results and discussion. Section V concludes the paper.

## II. LITERATURE REVIEW

This section reviews some previous studies on bone fractures to identify gaps in the existing literature.

Aneeza et al. [9] points to the fact that despite the progress in X-ray imaging technology and clinical expertise, manual bone fracture diagnosis remains a growing challenge. For this, they used deep architectures of DenseNet and VGG-19 CNNs to detect fractures from medical X-ray images using well structured and a diverse dataset. Proposed MobLG-Net (MobileNet-LGBM) based features resulted in the mild deviation of the LGBM and LR models, with the ones trained to predict bone fractures, showed overall accuracy of 99%. Nevertheless, the model under study has been neither focused on model interpretability, implementation, nor on dataset diversity.

In another recent study, the authors address the limitations of conventional radiologist based fracture diagnosis and the importance of rapid and accurate identification process in emergency or low resources settings [10]. The authors offer a deep learning framework based on a ResNet-50 model where a large, annotated X-ray dataset is combined with data enhancement technique and the framework is trained on this. For both detection and classification tasks on this model, accuracy was 94.3% and 91.7% respectively and for real time image processing in two seconds. Nevertheless, when the complexity of fracture and quality of images are high, the performance was bad for the challenging classification task, thus illustrating the need of a more diverse dataset for training and evaluation.

Challenges such as observer variability and shortages in the radiologist pools in clinical settings make bone fracture detection a critical need, as mentioned in this paper [11]. The ensemble deep learning framework, we propose an ensemble deep learning framework on ResNet-50, DenseNet-121, and EfficientNet-B3 and use of a weighted voting mechanism to improve the prediction accuracy. With limited training data, and across complex anatomical regions, the model performed with 0.977 AUC, 94.8% sensitivity, and 95.2% accuracy. The research is not scalable because it has lower accuracy at most subtle fractures and very high computational demands.

Another study discusses how the use of x-ray imaging and clinical infrastructure have not diminished the need for manual diagnosis of bone fractures [12]. In result they build an ensemble deep learning model which consists of MobileNet-V2, VGG-16, Inception-V3, ResNet-50, used for preprocessing, including histogram equalization and Global Average Pooling, learn on Mura-v1.1 dataset. This model achieved 92.96% accuracy, 91.62% recall, and a 92.14% F1 score, and was better than all the individual architectures at detecting humerus fractures. Despite this, its applicability to other fracture types and use of a public dataset hamper its generalization in the clinical settings.

In bone fracture identification, Nath et al. [13] work on the high misdiagnosis rate to classify X-ray images from the MURA dataset with fractured or non fractured bones using AfricanDNN model based on Deep Convolutional Neural Network (DCNN) model using AlexNet. It is an automatic preprocessing approach, top layer retraining, and localization method with bounding boxes. The best results, including 95% classification accuracy, yield better results than those obtained with SURF or MLP-based BPNN. However, the limits of the study's generalizability as a result of looking only at general classification and no performance for specific fracture types or anatomical regions are unclear.

Ahmed et al. [14] describe a machine learning approach for the detection and degradation of bone fractures in X-ray images. The paper underscores the significance of proper diagnosis in the field of medicine, especially with the ever rising incidence of bone fractures. Fourthly, the model designed is composed from four principal modules named preprocessing, edge detection, feature extraction and classification. A dataset of X-ray images were studied using different machine learning algorithms like Naïve Bayes, Decision Tree, Nearest Neighbors, Random Forest and Support Vector Machine (SVM). However, the proposed method in providing a better accuracy of 92.86% in fracture detection and classification of the SVM algorithm proves the efficiency of the proposed method.

## III. METHODOLOGY

This section presents the methodology for bone fracture detection. Figure 1 illustrates the workflow of the proposed model.

### A. Data Collection

In this study, we used respiratory related medical imaging dataset that was taken from publicly available repository on Kaggle. In total there are 9,463 samples, among them 4,840 are *fractured* and 4,623 are *not fractured* labeled samples. The Figure 2 below illustrates a sample from the dataset.

### B. Preprocessing

In this study, several preprocessing techniques were applied to enhance the detection of bone fractures in medical images.

*1) Image Resizing and Normalization:* In order to analyze, we had to standardize the pixels of the images to the size $224 \times 224$ pixels. Uniformizing input dimensions is necessary for deep learning models and is performed via resizing.

We also normalized the pixel values using the division of each pixel value by 255 to scale the pixel values on (0, 1). It helps to make the model better converge and better perform.

*2) CLAHE (Contrast Limited Adaptive Histogram Equalization):* With CLAHE subtler features such as fractures become more visible The image is divided into tiles, then histogram equalized, with a contrast limit preventing the amplification of noise [15].

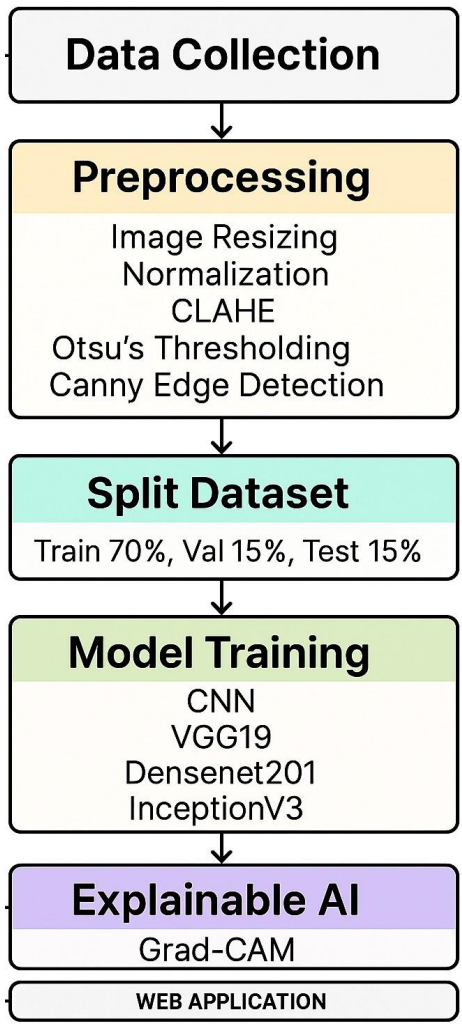

Fig. 1. Workflow Diagram of this study.

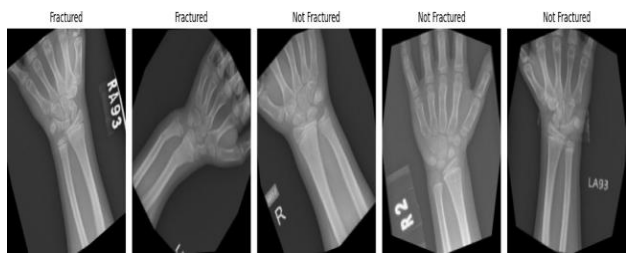

Fig. 2. Sample of fractured and non-fractured of Dataset.

*3) Otsu's Thresholding:* Through Otsu's technique the image gets automatically divided when the method finds the best threshold to split between the fractured bone and the surrounding area [16]. The threshold $T$ is found by minimizing the intraclass variance:

*4) Canny Edge Detection:* Canny edge detection uses its method to see fast changes in intensity that mark fracture edges [17].

In Figure 3, it can be seen that CLAHE improves contrast, Otsu's method automates segmentation, and Canny edge detection outlines fracture boundaries.

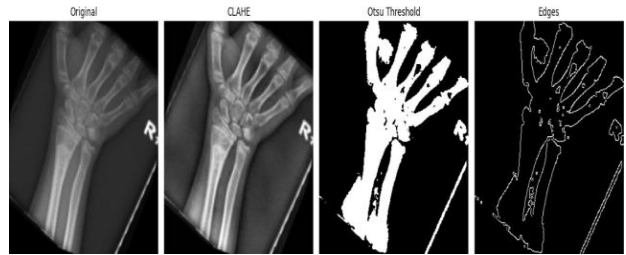

Fig. 3. Preprocessed image showing the effects of CLAHE, Otsu's Thresholding, and Canny Edge Detection.

### C. Dataset Splitting

A 70-15-15 split is applied to the dataset. After splitting, the label counts for the training, validation, and test datasets are shows in Table I:

TABLE I
LABEL COUNTS FOR TRAINING, VALIDATION, AND TEST DATASETS

| Dataset | Fractured | Not Fractured |
|---|---|---|
| Training | 3388 | 3236 |
| Validation | 726 | 693 |
| Test | 726 | 694 |

### D. Model Training

The models we trained on were Inception-V3, DenseNet-201, and a custom CNN. They were all used to detect bone fractures. Furthermore, we applied a Modified VGG-19 architecture, which we modify the classifier of it with dense (FC) layer and ReLU activation functions, to further improve the extracting and the classification performance. The performance of these models was optimized in terms of accuracy and were compared between each other. The architecture of the modified VGG-19 model is explained in Figure 4 below.

*1) Trainable and Non-Trainable Parameters:* The total number of parameters of each deep learning model used in this study is presented in Table II. Even though two modified VGG-19 and CNN architectures have much less parameters than Inception-V3 and DenseNet-201, it means that the computational complexity are much lower. The comparison gives an idea about model complexity and performance in fracture detection.

TABLE II
TOTAL PARAMETERS FOR EACH MODEL

| Model | Total Parameters |
|---|---|
| VGG-19 (Modified) | 23,195,656 |
| CNN (Modified) | 25,517,512 |
| Inception-V3 | 24,974,056 |
| DenseNet-201 | 30,144,008 |

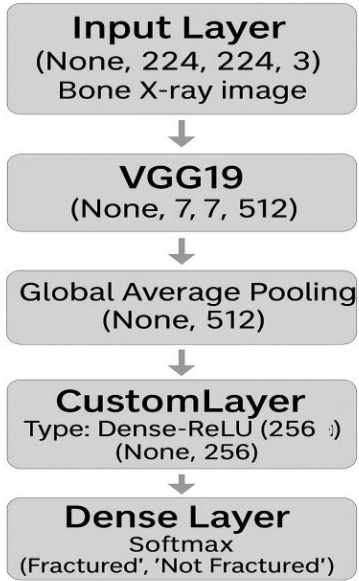

Fig. 4. Proposed VGG-19 Architecture.

*2) Experimental Setup:* A set of well-defined hyperparameters and configuration were used to ensure effective training and optimal performance of proposed fracture detection model. These are very carefully chosen so as to reduce learning efficiency while avoiding overfitting, but ensuring better generalization. A summary of the detailed training configuration is given in Table III.

TABLE III
MODEL TRAINING SETTINGS

| Parameter | Value |
|---|---|
| Optimizer | Adam |
| Learning Rate | 0.0005 |
| Loss Function | Sparse Categorical Cross-Entropy |
| Batch Size | 32 |
| Epochs | 40 |
| Early Stopping | Patience = 10, restore best weights |
| Input Image Size | 224 × 224 × 3 (RGB) |
| Base Model | VGG-19,CNN, Inception-V3, DenseNet-201 |
| Activation (Final Layer) | Softmax |

### E. Use of Explainable AI

To make the results more interpretable, we used Grad-CAM to visualize the areas of the image inputs which were important for a model's prediction.

### F. Web Application

Finally, the best model was deployed on Hugging Face Spaces using Gradio so real time fracture detection is possible via a user friendly web interface.

### G. Evaluation Metrics

The performance of the trained models is evaluated in terms of training and test accuracy, precision,recall F1-score, AUC, and both training and inference time.

## IV. RESULT AND DISCUSSION

This section discusses the results obtained from the applied model, highlighting its performance and effectiveness in bone fracture detection.

### A. Training and Validation Performance

In Table IV, the bone fracture detection model is applied and results are shown concerning models used in training and validation. The training accuracy achieved by the trained VGG-19 architecture of 99.95% and 98.45% validation accuracy is significantly higher that of the custom CNN trained architecture resulting in a 99.92% training accuracy and 98.31% validation accuracy. Inceptions-V3 has a training accuracy of 99.75% and validation accuracy of 97.96%, DenseNet-201 had a training accuracy of 99.91% and validation accuracy of 98.30%. This resulted in a good training accuracy and a good learned ability to handle fracture related features for all models.

Validation accuracy and loss curves of the best VGG-19 model converged and stabilized, as can be observed under Figure 5, which represented training convergence and stability of the best VGG-19 model.

Additionally, the training time, inference time and the memory usage for each model are shown in the table below in order to show the computational efficiency and resource consumption of these models.

TABLE IV
TRAINING AND VALIDATION ACCURACY OF DIFFERENT MODELS

| Model | Training Accuracy (%) | Validation Accuracy (%) |
|---|---|---|
| VGG-19 | 99.95 | 98.45 |
| Custom CNN | 99.92 | 98.31 |
| Inception-V3 | 99.75 | 97.96 |
| DenseNet-201 | 99.91 | 98.30 |

Computational efficiency is presented in table V. The Custom CNN achieved the fastest inference time (2.03s) but used the most memory (12118.82 MB) and because of that, it makes the Custom CNN suitable for real time apps. It appears that VGG-19 had a balanced profile with low training time (69.57 s), moderate inference time (7.89 s) and the least memory usage (7266.95 MB). Although they were accurate, the training and inference times of Inception-V3 and DenseNet-201 were significantly higher than other methods considered ahead. Specifically, DenseNet-201 was the hungriest of them with the highest training and inference times. Therefore, it is shown that Custom CNN and VGG-19 are the more appropriate choices for real time deployment, given limited resources. To further support this comparison, Table VI gives a summary of the testing performance of each model.

TABLE V
TRAINING TIME, INFERENCE TIME, AND MEMORY USAGE FOR DIFFERENT MODELS

| Model | Training Time (s) | Inference Time (s) | Memory Usage (MB) |
|---|---|---|---|
| VGG-19 | 69.57 | 7.89 | 7266.95 |
| Custom CNN | 71.04 | 2.03 | 12118.82 |
| Inception-V3 | 182.43 | 17.14 | 9799.43 |
| DenseNet-201 | 357.89 | 52.94 | 10858.00 |

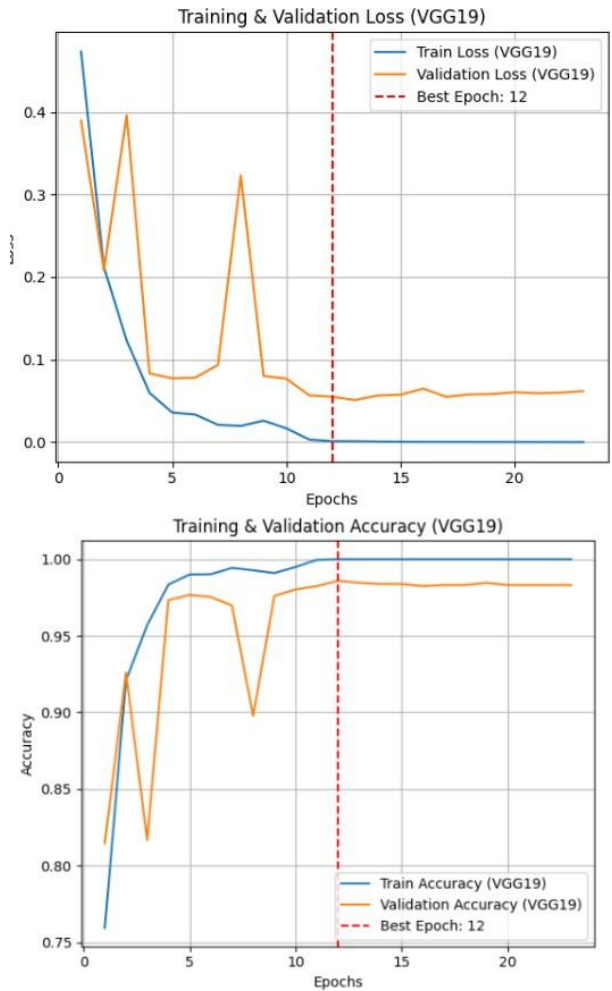

Fig. 5. Validation accuracy and loss curves for the VGG-19 model.

## B. Testing Performance

The testing accuracy as well as the AUC scores of the tested models are summarized in Table VI. Overall, the testing accuracy of VGG-19 reached the highest value at 99.78% with a perfect AUC score of 1.00, indicating a very good discriminatory power. The results of the Custom CNN were also satisfactory, achieving 99.01% accuracy and AUC of 0.9979, indicating good generalization, yet it is lightweight. In fact, DenseNet-201 was the closest with 99.37% accuracy and 0.9937 AUC; Inception-V3 performed relatively worse compared to others with 97.39% accuracy and 0.9963 AUC. These results confirm the VGG-19 robustness for correctly identifying fracture and non fracture cases. Finally, the confusion matrix and ROC curve of VGG-19 is shown in Fig 6 & 7, for further insights in its classification performance.

The confusion matrix for modified VGG-19 model to classify between fractures versus non fractures is shown in Figure 6. Using this, the model has yielded 692 true negatives, 725 true positives, 2 false positives, 1 false negative. The presentation of these results shows high accuracy with little error, and the matrix is visually highlighting correct predictions are dominating over incorrect ones. This confirms that the model is capable of discriminating between fractured and non fractured cases and its ability to do so is very accurate.

TABLE VI
TESTING ACCURACY AND AUC SCORE OF DIFFERENT MODELS

| Model | Testing Accuracy (%) | AUC Score |
|---|---|---|
| VGG-19 | 99.78 | 1.00 |
| Custom CNN | 99.01 | 0.9979 |
| Inception-V3 | 97.39 | 0.9963 |
| DenseNet-201 | 99.37 | 0.9937 |

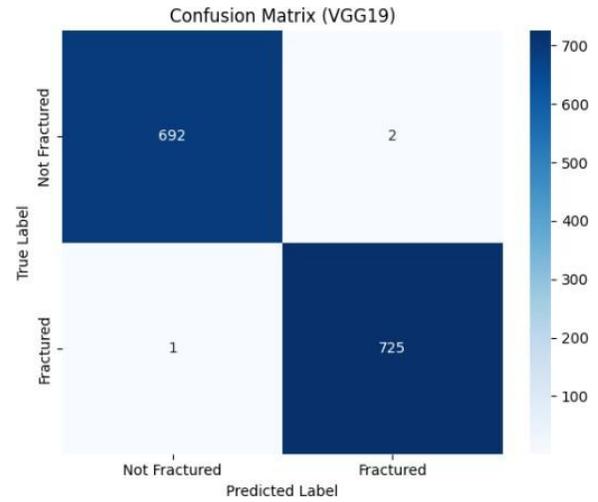

Fig. 6. Confusion matrix for the VGG-19 model.

The AUC of the VGG-19 model is 1.0 as shown in Figure 7, making it a perfect ROC curve. This demonstrates the ability of this model to wonderfully discriminate between fractured and non-fractured cases; not offering any trade off between sensitivity and specificity, meaning that this model is at his optimum.

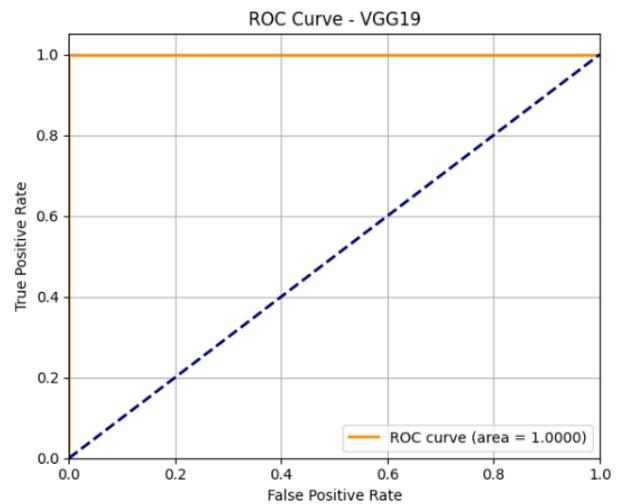

Fig. 7. ROC Curve for the VGG-19 model.

## C. VGG-19 Decision Making Process

The Grad-CAM visualizations of modified VGG-19 model's interpretability are shown in figure 8. For a fractured wrist, the model correctly identifies it in the top row and has strong activation around the fracture site. The model fails to correctly classify a non-fractured forearm of the bottom row with diffuse attention across the bone shaft, located in the bottom row. These visualizations show the model's ability to our attention on clinically important aspects and increasing transparency and trust.

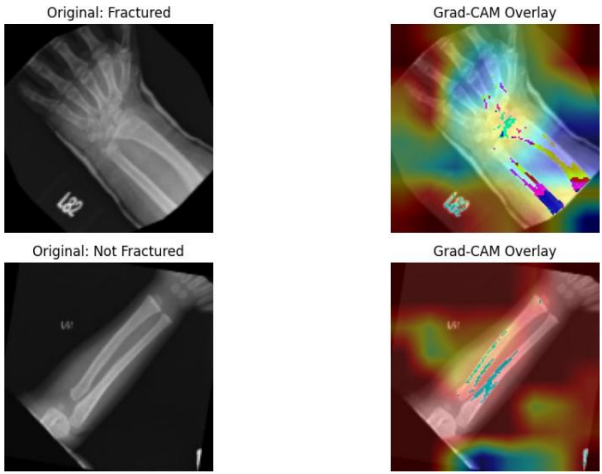

Fig. 8. Grad-CAM visualization of the modified VGG-19 model.

## D. Web Application

Furthermore, we built a real time fracture detection web app which we deploy on Hugging Face Spaces. Figure 9 shows the web app interface and its prediction results together with the prediction confidence. On one hand, this feature gives the user not only the fracture classification but also the confidence level of the prediction to increase model reliability and transparency while being used in clinical settings. Curiously, it can return the results in just 0.5 seconds, thereby allowing quick decisions on time critical medical matters.

## E. Comparative Analysis

This study's Modified VGG-19 model does better than all the compared study in VII, and tables show that it has a better accuracy than all of them that is the best among the claimed study. This study unique feature includes including Explainable AI (XAI) to boost model's transparent decision making process, and its superior accuracy. In addition, it facilitates real-time detection, a novel and helpful property when compared to the literature that is reviewed, as it streamlines the procedure of fracture detection.

## V. CONCLUSION

In this work we present a highly accurate and interpretable transfer learning based approach to detect bone fracture using Explainable AI (XAI) techniques combined with a modification of VGG-19 model. The proposed modified VGG-19 shows that it could serve as a reliable diagnostic tool with an accuracy of 99.78%, an AUC of 1.00, and potential in clinical settings. Additionally, its low parameter count and memory usage make it highly efficient, ensuring suitability for deployment in resource-constrained environments, where quick and accurate diagnostics are critical. Through this integration, we not only enforce trust of the model's prediction but give the healthcare professional deeper insights of fracture localization.

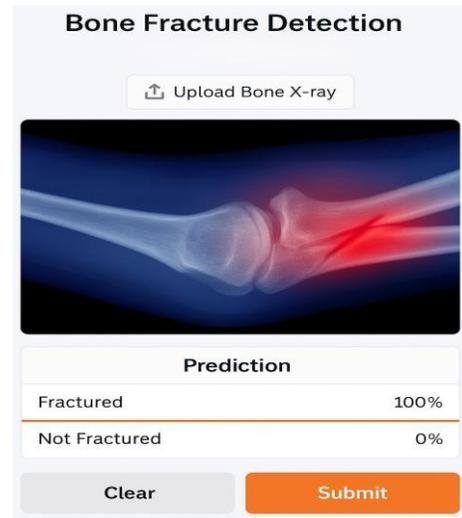

Fig. 9. Web Application for real time Bone fracture detection.

TABLE VII
COMPARATIVE ANALYSIS OF VARIOUS FRACTURE DETECTION MODELS.

| Study | Model | Accuracy | XAI | Web App |
|---|---|---|---|---|
| [9] | VGG-19 | 99% | No | No |
| [11] | EfficientNet-B3 | 95.2% | No | No |
| [12] | MobileNet-V2 | 92.96% | No | No |
| [13] | AlexNet | 95% | No | No |
| [14] | SVM | 92.86% | No | No |
| This Study | Modified VGG-19 | 99.78% | Yes | Yes |

Additionally, the development of a real time and user friendly web application bridges the gap between the current development in the field of advanced AI and their use in healthcare in real world with a quick and available diagnosis in various clinical setups. This represents a major step forward in a union of performance, interpretability, and usability of AI enabled medical diagnostics.

The dataset is expanded to include more X-ray image diversity in future work. More performance boost may be attained using alternative deep learning architectures as well as with advanced AI such as self-supervised or federated learning. It will also seek to improve usability and integration into clinical system to make the tool practical and functional in clinical setting of the real world.